\documentclass[12pt]{article}
\pdfoutput=1

\usepackage{amsmath,amssymb,amscd}
\usepackage{listings}
\usepackage{caption}
\usepackage{dsfont}
\usepackage{slashed}
\usepackage{color}
\usepackage{ulem}

\usepackage[pdftex]{graphicx}
\usepackage{epstopdf}
\usepackage{subfigure}
\usepackage{epsfig}
\usepackage{listings}
\usepackage{caption}
\usepackage{cite}

\usepackage{multirow}

\newcommand{\beq}{\begin{equation}}
\newcommand{\eeq}{\end{equation}}
\newcommand{\nbea}{\begin{align*}}
\newcommand{\neea}{\end{align*}}
\newcommand{\nbeq}{\begin{equation*}}
\newcommand{\neeq}{\end{equation*}}

 \usepackage{multirow}
\usepackage{array}
\newcolumntype{M}[1]{>{\centering\arraybackslash}m{#1}}
\newcolumntype{N}{@{}m{0pt}@{}}

\begin{document}


\pagestyle{empty}

\baselineskip=21pt
\rightline{{\fontsize{0.40cm}{5.5cm}\selectfont{KCL-PH-TH/2016-06, LCTS/2016-04, CERN-TH/2016-032}}}
\rightline{{\fontsize{0.40cm}{5.5cm}\selectfont{ACT-02-16, MI-TH-1608}}}
\vskip 0.75in

\begin{center}

{\large {\bf Remarks on Graviton Propagation in Light of GW150914}}

\vskip 0.5in

 {\bf John~Ellis}$^{1,2}$,~
   {\bf Nick~E.~Mavromatos}$^{1,2}$
and {\bf Dimitri~V.~Nanopoulos}$^{3}$

\vskip 0.5in

{\small {\it

$^1${Theoretical Particle Physics and Cosmology Group, Physics Department, \\
King's College London, London WC2R 2LS, UK}\\
\vspace{0.25cm}
$^2${Theoretical Physics Department, CERN, CH-1211 Geneva 23, Switzerland}\\
\vspace{0.25cm}
$^3${George P. and Cynthia W. Mitchell Institute for Fundamental Physics and Astronomy, Texas A\&M University, College Station, TX 77843, USA;\\
\vspace{0.25cm}
Astroparticle Physics Group, Houston Advanced Research Center (HARC), Mitchell Campus, Woodlands, TX 77381, USA; \\
Academy of Athens, Division of Natural Sciences, Athens 10679, Greece}
}}

\vskip 0.5in

{\bf Abstract}

\end{center}

\baselineskip=18pt \noindent


{\small
The observation of gravitational waves from the Laser Interferometer Gravitational-Wave Observatory (LIGO) event 
GW150914 may be used to constrain the possibility of Lorentz violation in graviton propagation, and
the observation by the Fermi Gamma-Ray Burst Monitor of a transient source in apparent coincidence may be used to
constrain the difference between the velocities of light and gravitational waves: $c_g - c_\gamma < 10^{-17} $.}


\vskip 0.75in

\leftline{ {February 2016}}

\newpage
\pagestyle{plain}

The discovery of gravitational waves by the Laser Interferometer Gravitational-Wave Observatory (LIGO) in event 
GW150914~\cite{LIGO} opens a new era in astronomy, making possible the measurement of astrophysical
processes that have been inaccessible to observations with electromagnetic waves. The question then arises
what fundamental physics we can learn from gravitational wave observations in general and LIGO event 
GW150914 in particular. As examples, the LIGO Collaboration itself~\cite{LIGO2} has reported an upper limit on the
graviton mass $m_g < 10^{-22}$~eV, and it has been suggested that observations of binary black-hole mergers
could constrain models of quantum physics near black-hole event horizons~\cite{Giddings}.

In this comment we derive two additional constraints on graviton propagation, assuming that it is massless. 
First, the LIGO data on GW150914
can be used to constrain the possibility of Lorentz violation~\cite{LIV} in gravitational wave propagation,
assuming that low-frequency gravitational waves (low-energy gravitons) travel at the conventional speed of light {\it in vacuo}
$c$ (that we set to unity from now on), whereas higher-frequency waves (higher-energy gravitons)
may travel at frequency- (energy-)dependent velocities. Secondly, assuming instead that the velocities of gravitational
and electromagnetic waves $c_g$ and $c_\gamma$ are
frequency- (energy-)independent, we use the apparent coincidence
of a transient source with photon energies $> 50$~keV observed by the Fermi Gamma-Ray Burst Monitor (GBM)~\cite{fermigbm}
to constrain the difference between the velocities of light and gravitational waves {\it in vacuo}: $c_\gamma - c_g < 10^{-17} c$.

The LIGO constraint on the graviton mass was obtained from a detailed numerical comparison of the
measured GW150914 wave-form with that calculated for a black-hole merger~\cite{LIGO2}. We recall that the GW150914 signal
consisted of a `chirp' of increasing frequencies $\omega \sim 100$~Hz,
with a range of frequencies $\Delta \omega = {\cal O}(100)$~Hz.
The presence of a gravitino mass would induce an energy- (frequency-)dependent deviation of the
velocities of the waves emitted during the `chirp' from that of light: $\Delta v|_{m_g} \simeq - m_g^2/2 \omega^2$.
Such a deviation $\Delta v$ would cause a dispersion in their arrival times~\cite{will}, which is constrained by
concordance of the observed signal with numerical relativity calculations.

It was suggested in~\cite{aemn} that quantum-gravitational effects might induce an energy- (frequency-)dependent velocity
of propagation {\it in vacuo} for both electromagnetic and gravitational waves $\Delta v|_{LVn} \simeq - \xi (\omega/M_n)^n: n = 1$ or 2
where $M_n$ is some large mass scale, where $\xi = +1 (-1) $ for subluminal (superluminal) propagation and low-energy (-frequency)
waves would travel at the conventional velocity of light. 
Such a Lorentz-violating  effect would give rise to an energy-dependent dispersion in the
arrival times of gravitational waves, though with a different energy dependence from a graviton mass.
Such Lorentz violation might be induced by the effects of space-time foam on wave
propagation, in which case one might expect that $M_n = {\cal O}(M_P) \sim 10^{19}$~GeV.
We recall that subluminal propagation is implied by
concrete models of space-time foam within brane theory~\cite{delays}.

It was suggested~\cite{aemns} that the existence of such Lorentz violation~\cite{LIV} could 
best be probed by studying energetic photon
emissions from distant transient astrophysical sources such as gamma-ray bursters (GRBs) or active
galactic nuclei (AGNs). The most sensitive limits on such an effect have been placed by MAGIC~\cite{MAGIC} and HESS~\cite{HESS}
observations of AGNs and Fermi observations of GRBs~\cite{GRB2}. No analogous constraint has 
previously been established on the possibility of such Lorentz violation in the propagation of gravitational waves
(gravitons), but this is now possible with the LIGO discovery of the gravitational waves produced in event
GW150914, as we now discuss.

In order to obtain a first order-of-magnitude constraint on Lorentz violation in gravitational wave propagation,
we assume that a detailed numerical analysis could establish a limit on $\Delta v|_{LVn}$ as the LIGO
Collaboration established on $\Delta v|_{m_g}$, considering as an illustration the linear subluminal LV case $n = 1$.
\begin{equation}
\Delta v|_{LV1} \; = \;  - \left( \frac{\omega}{M_1} \right) \; \simeq \; \Delta v|_{m_g} = - \left( \frac{m_g^2}{2 \omega^2} \right) \, .
\label{dv}
\end{equation}
Accordingly, we estimate $M_1 \gtrsim 2 \omega^3/m_g^2$, where we estimate $\omega \simeq 100$~Hz and
use the LIGO limit $m_g \lesssim 10^{-22}$~eV to obtain~\footnote{Since the source of event GW150914
is estimated to have a redshift $z \lesssim 0.1$, effects due to cosmological expansion and uncertainties in the cosmological model do not affect significantly our estimate. We expect a similar limit in the superluminal case where the LV effect would have the opposite sign
from a graviton mass.}
\begin{equation}\label{m1}
M_1 \; \gtrsim \; 100 \; {\rm keV} \, . 
\end{equation}
Clearly, this is many orders
of magnitude less than the limit $M_1 \gtrsim 10^{19}$~GeV obtained for photons (electromagnetic waves)~\cite{MAGIC, HESS, GRB2},
and also many orders of magnitude less than the naive expectation based on ideas of space-time foam,
but it is a start~\footnote{We note that substituting the lower limit (\ref{m1}) back into the expression (\ref{dv}) for
gravitational wave with $\omega \sim 100 Hz$ yields (in our subluminal example): $ \Delta v|_{LV1} = c_g - 1 \sim 4 \times 10^{-18} $,
which corresponds also to the deviation of the propagation speed of massive gravitons from the massless case in a Lorentz-invariant vacuum.}. 

The limit (\ref{m1}) corresponds to a variation in in the velocities of the ${\cal O}(100)$~Hz gravitational waves emitted by GW150914
at the level of ${\cal O}(10^{-17})$. As such, it would have a negligible effect on the physical scale of the merger event, and hence on
the waveform of the gravitational waves ($10^{-12}$~m on a scale of 100~km). Certainly, no such effect would be expected in the
framework~\cite{aemn, delays} that motivated this study.
The sensitivity to Lorentz violation should be evaluated using a full numerical simulation of
the black-hole merger, and one might expect the sensitivity to increase by an order of magnitude with the observation of a
neutron-star merger with characteristic frequency $\omega \sim 1000$~Hz. Looking to the future, observations
of the mergers of more massive black holes would not give an increase in sensitivity, but a new frontier
would be opened if/when tensor CMB perturbations are observed~\cite{CMBLV}.

We now discard the possibility of energy (frequency) dependence in the velocities of gravitational and 
electromagnetic wave propagation, and ask instead how similar these velocities must be.
For this analysis~\footnote{For previous discussions of the possibility of such an analysis and the assumptions involved, see~\cite{previous}.}, 
we use the arrival times of the
GW150914 signal and the apparently coincident flash of photons with energies $> 50$~keV observed
$\sim 0.4$~s later by the Fermi GBM~\cite{fermigbm}~\footnote{We note, however, that the significance of this
apparent coincidence is not high, and that the INTEGRAL
experiment did not see a signal of similar strength at similar energies~\cite{INTEGRAL}.}. The plausibility of the Fermi GBM signal
has been questioned~\cite{Lyutikov}, but also a number of models have been proposed to explain it~\cite{models},
which predict that any such photons would have been emitted in the aftermath of the merger.
Using the distance estimate of $\sim 10^9$~light-years
to the source of GW150914, the following {\it upper bound} on the difference between the
velocities of light and gravitational waves, $c_{g, \gamma}$:
\begin{equation*}
c_g - c_\gamma \; \lesssim \; 10^{-17} \, .
\end{equation*}
We note that the possibility $c_g < c_\gamma$ cannot be excluded if the photons were emitted more than 0.4~s after
the gravitational waves.

Another constraint on the velocity of gravitation waves from GW150914 was given in~\cite{Blas}, but this is much more
stringent~\footnote{Previous upper limits on Lorentz violation in photon propagation~\cite{MAGIC, HESS, GRB2}
ensure that the $c_\gamma$ inferred from
the putative Fermi GBM observation of GW150914 can be identified with the standard velocity of light with an
accuracy much greater than this constraint.}.  For completeness, we recall that an indirect {\it lower bound} on $c_g$ was
set in~\cite{moore}, derived from the non-observation of gravitational Cherenkov radiation from high-energy cosmic rays.
If the origin of the latter is assumed to be extragalactic then $c_\gamma - c_g < 2 \times 10^{-19}\, c_\gamma$, 
otherwise the bound is weaker: $c_\gamma - c_g < 2 \times 10^{-15}\, c_\gamma$.
On the other hand, we emphasise that our constraint $c_g - c_\gamma \lesssim 10^{-17}$ cannot be regarded as
definitive, since the Fermi GBM report~\cite{fermigbm} cannot be regarded as a definitive observation of a photon flash in
coincidence with GW150914. We await with interest possible future observations of light flashes coincident
with gravitational wave events.

\section*{Note Added}

The LIGO Collaboration has recently reported~\cite{LIGObox} the observation of a second gravitational wave event, GW151226,
interpreted as the merger of two black holes with masses $\sim 14.2$ and 7.5 solar masses at a distance
$d \sim 440$~Mpc. The peak amplitude of the gravitational wave train is at $\sim 450$~Hz. We make a rough
estimate of the sensitivity to $M_1$, as follows. In view of the consistency of the gravitational wave train with
calculations in general relativity, we assume that
\begin{equation}
\Delta v \cdot d \; \lesssim \; \frac{1}{\omega} \, ,
\label{LIGO2}
\end{equation}
where $\Delta v$ is the fractional deviation of the wave propagation velocity from that of light,
and we take (conservatively) $\omega \sim 200$~Hz, leading to
\begin{equation}
M_1 \; = \; \frac{\omega}{\Delta v} \; \gtrsim \; 400~{\rm keV} \, .
\label{newM1}
\end{equation}
This very crude estimate would need to be refined by a detailed numerical analysis, but it reinforces the point
that mergers of smaller objects can give more stringent constraints on Lorentz violation.

To date, there have been only negative results from searches~\cite{nogamma} for a possible electromagnetic counterpart to
GW151226, so it provides no further constraint on any possible possible difference between the velocities of gravitational
and electromagnetic waves.

We note finally that the LIGO Collaboration describe also~\cite{LIGORun1BH}
a possible signal (LVT151012) for a merger of a pair of black holes with
masses $\sim 28, 16$ solar masses at a distance $\sim 1000$~Mpc. If real, this merger would have a sensitivity to $M_1$ intermediate
between those of GW150914 (\ref{m1}) and GW151226 (\ref{newM1}).

\section*{Acknowledgements}

The research of JE and NEM was supported partly by the London Centre for Terauniverse Studies (LCTS), 
using funding from the European Research Council via the Advanced Investigator Grant 26732, 
and partly by the STFC Grant ST/L000326/1. JE thanks the Universidad de Antioquia, Medell\'in, for its
hospitality, using grant FP44842-035-2015 from Colciencias (Colombia), and he thanks Antonio~Enea~Romano and Diego~Restrepo
for discussions. The research of DVN
was supported in part by the DOE grant DE-FG02-13ER42020.

\end{document}